\documentclass[acmtog]{acmart}

\usepackage{multirow}
\usepackage{fancyhdr}
\usepackage{graphicx}
\usepackage{bbding}
\usepackage{verbatim}
\usepackage{booktabs}
\pdfpageattr {/Group << /S /Transparency /I true /CS /DeviceRGB>>}

\AtBeginDocument{%
  \providecommand\BibTeX{{%
    \normalfont B\kern-0.5em{\scshape i\kern-0.25em b}\kern-0.8em\TeX}}}
\acmConference[SIGGRPAH '22]{SIGGRAPH '22}{August 07--11, 2022}{Vancouver}

\copyrightyear{2022}
\acmYear{2022}
\setcopyright{acmcopyright}
\acmConference[SIGGRAPH '22 Conference Proceedings]{Special Interest Group on Computer Graphics and Interactive Techniques Conference Proceedings}{August 7--11, 2022}{Vancouver, BC, Canada}
\acmBooktitle{Special Interest Group on Computer Graphics and Interactive Techniques Conference Proceedings (SIGGRAPH '22 Conference Proceedings), August 7--11, 2022, Vancouver, BC, Canada}
\acmPrice{15.00}
\acmDOI{10.1145/3528233.3530709}
\acmISBN{978-1-4503-9337-9/22/08}

\acmSubmissionID{papers\_316}

\definecolor{orchid}{rgb}{0.85, 0.44, 0.84}
\definecolor{MyGreen}{RGB}{70,189,96}

\newcommand{\cmark}{\color{MyGreen} \CheckmarkBold}%
\newcommand{\xmark}{\color{red} \XSolid}%

\newcommand{\tabincell}[2]{\begin{tabular}{@{}#1@{}}#2\end{tabular}}

\newcommand{\dinesh}[1]{{{}}}

\begin{document}

\fancyhead[ER]{\sffamily \footnotesize Xiaoyu Pan, Jiaming Mai, Xinwei Jiang, Dongxue Tang, Jingxiang Li,\\ Tianjia Shao, Kun Zhou, Xiaogang Jin, and Dinesh Manocha}

\title[Predicting Loose-Fitting Garment Deformations]{Predicting Loose-Fitting Garment Deformations Using Bone-Driven Motion Networks}

\author{Xiaoyu Pan}
\email{panxiaoyu6@gmail.com}

\author{Jiaming Mai}
\email{maijmwq@126.com}

\affiliation{%
  \institution{State Key Lab of CAD\&CG, Zhejiang University; ZJU-Tencent Game and Intelligent Graphics Innovation Technology Joint Lab}
  \city{Hangzhou}
  \country{China}}

\author{Xinwei Jiang}
\email{wesleyjiang@tencent.com}

\author{Dongxue Tang}
\email{julytang@tencent.com}

\author{Jingxiang Li}
\email{jingxiangli@tencent.com}

\affiliation{%
  \institution{Tencent NExT Studios}
  \city{Shanghai}
  \country{China}}

\author{Tianjia Shao}
\email{tianjiashao@gmail.com}

\author{Kun Zhou}
\email{kunzhou@acm.org}

\affiliation{%
  \institution{State Key Lab of CAD\&CG, Zhejiang University}
  \city{Hangzhou}
  \country{China}}

\author{Xiaogang Jin}
\authornote{Corresponding author.}
\affiliation{%
  \institution{State Key Lab of CAD\&CG, Zhejiang University; ZJU-Tencent Game and Intelligent Graphics Innovation Technology Joint Lab}
  \city{Hangzhou}
  \country{China}}
\email{jin@cad.zju.edu.cn}

\author{Dinesh Manocha}
\affiliation{%
  \institution{University of Maryland}
  \city{College Park}
  \country{USA}}
\email{dmanocha@umd.edu}

\begin{abstract}



  We present a learning algorithm that uses bone-driven motion networks to predict the deformation of loose-fitting garment meshes at interactive rates.  Given a garment, we generate a simulation database and extract virtual bones from simulated mesh sequences using skin decomposition. At runtime,  we separately compute low- and high-frequency deformations in a sequential manner. The low-frequency deformations are predicted by transferring body motions to virtual bones' motions, and the high-frequency deformations are estimated leveraging the global information of virtual bones' motions and local information extracted from low-frequency meshes. In addition, our method can estimate garment deformations caused by variations of the simulation parameters (e.g., fabric's bending stiffness) using an RBF kernel ensembling trained networks for different sets of simulation parameters. Through extensive comparisons, we show that our method outperforms state-of-the-art methods in terms of prediction accuracy of mesh deformations by about 20\% in RMSE and 10\% in Hausdorff distance and STED. The code and data are available at \textcolor{magenta}{\textit{\url{https://github.com/non-void/VirtualBones}}}.
\end{abstract}

\ccsdesc[500]{Computing methodologies~Animation}
\ccsdesc[500]{Computing methodologies~Machine learning}

\keywords{cloth animation, deep learning, skinning decomposition}

\begin{teaserfigure}
  \includegraphics[width=\textwidth]{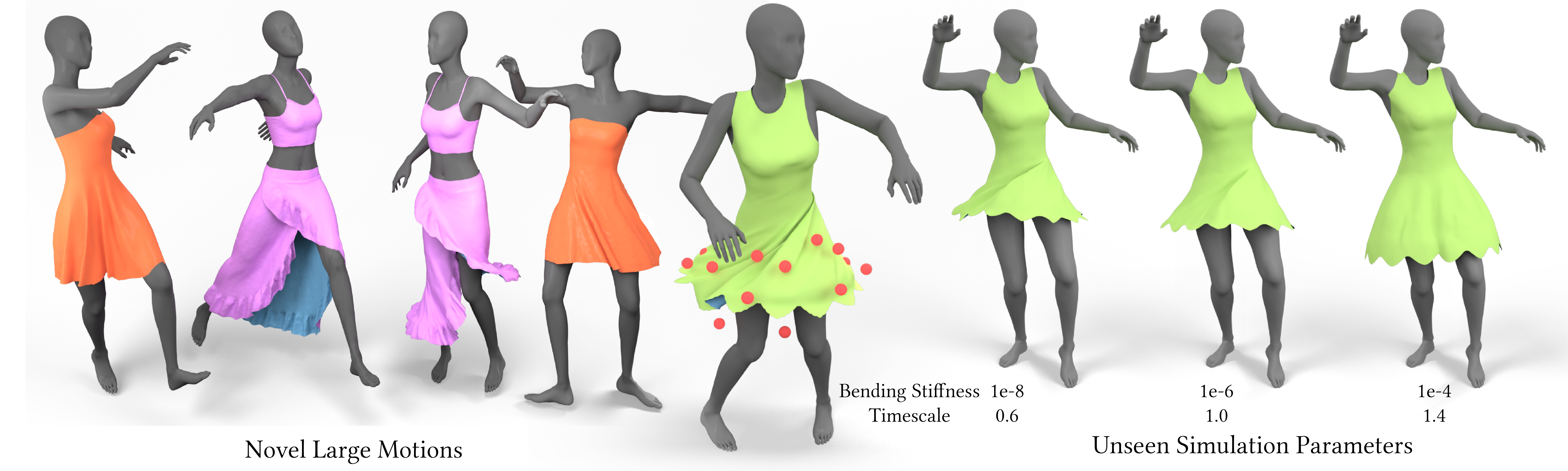}
  \caption{Given the body motion sequence and simulation parameters, our approach predicts the deformations of loose-fitting garments effectively. We transfer the body motion sequence to generate motion of garment's bones (red balls shown in the middle figure on the lower part of the garment). We use these bone-driven motion networks to predict large-scale  deformations caused by complex motions, such as flying, swirling and dropping on dresses (left). Our method can also be generalized to simulate dresses under unseen simulation parameters at an interactive rate (right).}
  \label{fig:teaser}
\end{teaserfigure}

\maketitle

\section{Introduction}



Garments, including loose-fitting garments with complex deformations, are critical for dressing humans. For dancing characters, the dynamic motion of their dresses, such as flying, dropping, swirling, or swinging, not only help express emotion~\cite{emotion1}, but also present rich characteristic details. Many interactive applications, such as games and VR, need the capability to dress characters or digital avatars with such loose-fitting garments.

Many data-driven or learning-based methods have been proposed for predicting cloth deformations. However, many of these methods~\cite{patel20tailornet,deepwrinkles,jin2020pixel,santesteban19,santesteban21,3dpeople,FullyConvolutionalGNN,cloth3d,garnet,bertiche2020pbns,deepsd} either focus on predicting garment shapes from static poses, or dynamics of tight-fitting garments, with deformations that closely follow the body. GarNet~\cite{garnet,garnet++} presents a method that retains the geometry features on garments by adopting curvature losses. Almost all deep-learning based methods demand a large amount of data, and PBNS~\cite{bertiche2020pbns} introduces a neural simulator trained by unsupervised learning to alleviate this issue. Few methods~\cite{Intrinsic, zhang2021dynamic} predict the deformations of loose-fitting garments, and even these methods may not be robust for large motions, or may not model high-frequency deformations~\cite{zhang2021dynamic}.
Moreover, current learning-based methods do not account for adjustments to simulation parameters (e.g., fabric's bending stiffness and simulator's timescale~\cite{HoudiniVellum}) in an interactive manner. They either infer simulation parameters~\textit{implicitly} given artist edited mesh~\cite{Intrinsic}, or adjust them by modifying weights of physical loss terms in the network~\cite{bertiche2020pbns}.


To simulate loose-fitting garments using deep learning, there are two main issues: complex deformations of loose-fitting garments and the heterogeneity between simulation parameters and body motions. In particular, these deformations may not follow the body shape closely, as is the case for tight-fitting garments. Many prior methods~\cite{patel20tailornet,santesteban19,santesteban21}, which deform garments according to their associated bodies' skinning functions, may result in artifacts.
Many image-based methods, such as geometry images~\cite{3dpeople}, displacement maps~\cite{jin2020pixel}, and normal maps~\cite{deepwrinkles}, are not suitable to represent such complex deformations, as they are mostly limited to tight-fitting clothes. Other learning methods use 3D mesh representations~\cite{Intrinsic,patel20tailornet,santesteban19,santesteban21}, but are unable to handle complex deformations for loose-fitting garments robustly (see Fig.~\ref{Qualitative-low-freq}). Moreover, the body motions and simulation parameters play distinctly different roles for such garments. Therefore, it is hard for a single network to learn them jointly. We need better network architectures that can simultaneously learn the two inputs.

\paragraph{Main Results:} We present a learning-based method to predict the 3D deformation of loose-fitting garments from body motion sequences and cloth simulation parameters automatically. A key aspect of our approach is extracting the virtual bones of the garment from a pre-computed simulation sequence and using them to drive the motion networks.
These virtual bones are used as an intermediate representation and to divide the prediction task into two sub-tasks: low- and high-frequency deformation learning.

As a preprocess, we generate physically-based simulations of loose-fitting garments using Houdini Vellum~\cite{HoudiniVellum}, and decompose them into low-frequency and high-frequency deformations. We then create a set of virtual bones from the simulation sequence by using skinning decomposition~\cite{BinhLe:TOG:2012}. At runtime, our model first learns to transfer the body motions to the virtual bones' transformations to estimate the garment's low-frequency deformations. We treat the virtual bones' motions as global information and also have the local information extracted by graph neural networks~\cite{EdgeConv,HeterSkinNet} from low-frequency mesh to generate high-frequency deformations. To handle heterogeneity between simulation parameters and body motions, we model deformations caused by simulation parameters and body poses differently. We train multiple networks that take body motions as inputs, and ensemble them using an RBF kernel to model simulation parameters.  The novel aspects of our work include:

\begin{itemize}
  \item The first deep-learning-based method for simulating the complex deformations of loose-fitting garments using virtual bones to efficiently represent low-frequency deformations and infer high-frequency deformations.
  \item A high-frequency deformation estimator that uses local information of the low-frequency mesh and global information of virtual bones' motions.
  \item A novel method to handle the heterogeneity between simulation parameters and body motions by modeling them in different types of networks.
\end{itemize}
We compare and highlight the benefits of our approach  with state-of-the-art methods~\cite{patel20tailornet,santesteban19,chen2021deep,zhang2021dynamic}.
Quantitatively,  our method achieves an improvement of 20\% in RMSE and 10\% in Hausdorff Distance and STED compared to the strongest baseline. Our qualitative results show that our method can effectively estimate the garments' dynamics with complex deformations, and that it outperforms competitive methods in terms of generating plausible deformations.

\section{Related Work}



\paragraph{Garment Animation and Deformation}

Garment animation and simulation is a well-studied problem in computer graphics. Many approaches have been developed and they can be broadly divided into two categories: physically-based simulations (PBS) and data-driven models. A key issue is the balance between computational speed and accuracy. In general, PBS methods generate high-quality results but tend to have high computational costs. In contrast,  data-driven models are less computationally costly but it is difficult for them to give any guarantee on accuracy. PBS methods tend to  model the real-world physics based on material properties of garments and deform them according to laws of physics using time integration, collision detection, and response computation~\cite{AAR,ASF,AirMesh,pbd1,Pcloth,pbd2,icloth,wang21}.

Data-driven models learn from a set of ground-truth garments, which are obtained from PBS or digital scanning. These methods are complementary to PBS methods, and their main benefit is faster performance, as highlighted in many recent papers \cite{patel20tailornet,santesteban19}. Most of recent mesh-based methods~\cite{patel20tailornet,deepwrinkles,jin2020pixel,santesteban19,santesteban21,3dpeople,FullyConvolutionalGNN,cloth3d,garnet,bertiche2020pbns,deepsd, habermann2021real, yu2018doublefusion} model tight-fitting clothes such as T-shirts and pants. \cite{santesteban19} model garment deformations with dynamic wrinkles using a recurrent neural network. To avoid smoothing out high-frequency details, TailorNet~\cite{patel20tailornet} disentangles garment deformations under static poses into low-frequency and high-frequency parts and models them separately. GarNet~\cite{garnet,garnet++} presents a method that retains the geometry features on garments by adopting curvature losses. Almost all deep-learning based methods demand a large amount of data, and PBNS~\cite{bertiche2020pbns} introduces a neural simulator trained by unsupervised learning to alleviate this issue. Some techniques~\cite{NearExhaustive, Intrinsic,zhang2021dynamic} have been proposed for loose-fitting garments with complex deformations that do not closely follow the body. \cite{NearExhaustive} use data-driven motion graphs to exhaustively model dynamic deformations of loose-fitting garments. In contrast, our learning‐based approach can be generalized to unseen motions and simulation parameters. \cite{Intrinsic} learn an intrinsic garment space that allows shape authoring, which projects the mesh to the latent space and \textit{implicitly} infers the latent representation of simulation parameters using the garment latent and the motion signature. However, their method may not model high-frequency deformations. \cite{zhang2021dynamic} adopt neural rendering to synthesize details onto images to retain these fine details. Our approach is complimentary to these methods and focuses on representing dynamic loose-fitting garments using virtual bones. Another line of works~\cite{bozic2021neuraldeformationgraphs, tiwari2021neural, saito2021scanimate, POP:ICCV:2021,Ma:CVPR:2021} uses neural fields to represent clothed human. Their outputs are unified neural fields, where human and cloth are inseparable and only one set of garments is associated with the character. In contrast, our method explicitly and directly models garments' meshes. Moreover, we can dress each character with different sets of garments.


\paragraph{Rigging}

\begin{figure*}[htbp]
  \centering
  \includegraphics[width=\linewidth]{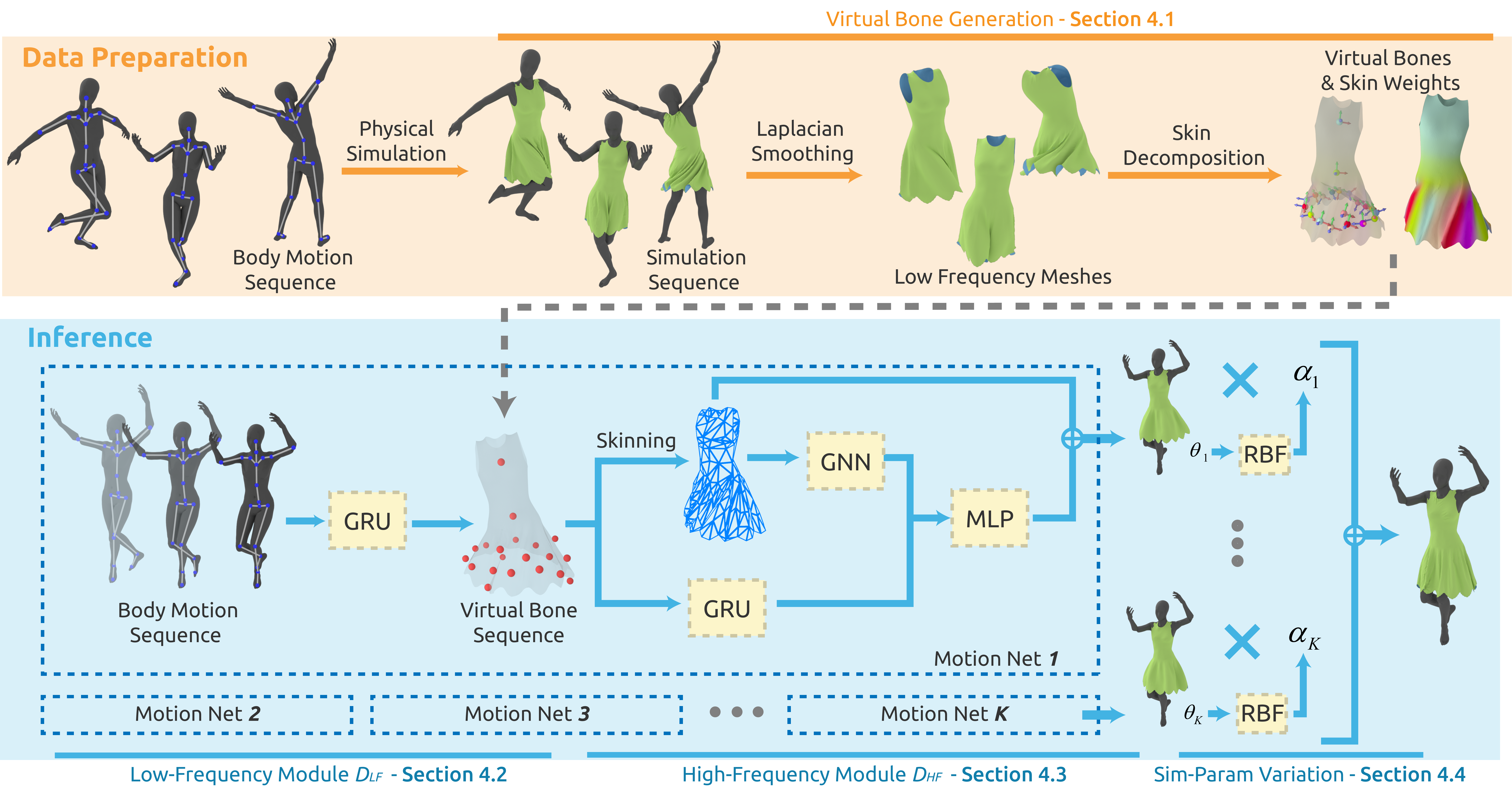}
  \caption{The top figure illustrates the generation of virtual bones and corresponding skin weights, which are obtained by performing skin decomposition on smoothed simulation sequences. The bottom figure demonstrates our inference process. The deformations caused by motions are modeled using $K$ pivot motion networks, which correspond to different sets of simulation parameters and take body motions as inputs. A motion network is composed of a low-frequency module and a high-frequency module. The former transfers the body motions to virtual bones' motions to predict the low-frequency deformations, and the latter estimates high-frequency deformations leveraging local information of the mesh and global information of virtual bones' motions. In terms of variations of simulation parameters, we obtain the result by combining the results of pivot motion networks using weights inferred by an RBF kernel.}
  \label{pipeline}
\end{figure*}

\begin{figure}[htbp]
  \centering
  \includegraphics[width=\linewidth]{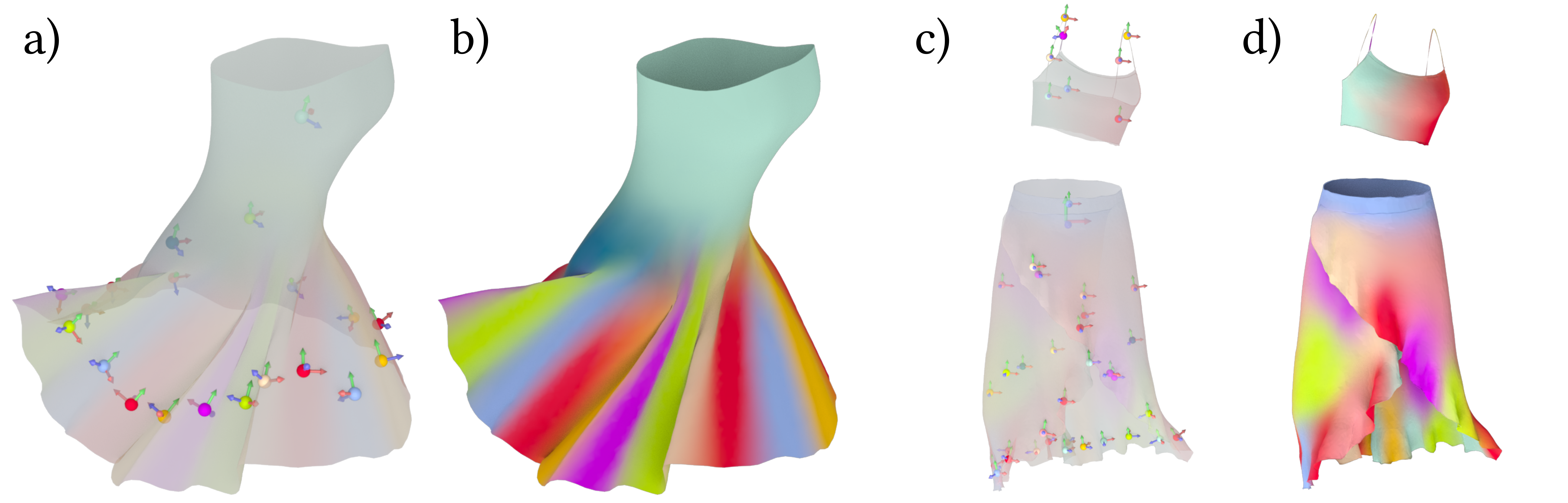}
  \caption{The virtual bones and corresponding skin weights, where the former are represented by balls and the latter are indicated by the shades on the mesh. (a), (b) 20 virtual bones and skin weights of Dress 3.  (c), (d) 40 virtual bones and skin weights of Dress 2.}
  \label{VirtualBoneWeights}
\end{figure}

Rigging is currently the predominant approach for animating characters, where users animate the mesh by controlling its underlying skeleton. Among these methods, linear blend skinning (LBS)~\cite{LBS} is the most popular algorithm because of its simplicity and efficiency. However, the representation space of LBS is considered to be limited to objects with rigid transformations. To represent deformable objects with rigging, researchers have developed techniques to expand the expressive power of LBS by enriching the space of skinning weights or using more transformations~\cite{kavan2009automatic,efficientskin}. Another line of work tries to augment skeleton-based rigs with physics, using coarse finite element simulations~\cite{capell2002interactive,mcadams}.
The former usage of rigging can be categorized into animation editing~\cite{LBS,kavan2009automatic,efficientskin,HeterSkinNet}, and joint extraction~\cite{BinhLe:TOG:2012}. We also use LBS in our approach.

\section{Overview}


Given the body motions $M_{b}^{\{1,...,t\}}$, i.e., the rotations of joints and translations of the body, and the set of simulation parameters $\boldsymbol{\theta}$, e.g., fabric's bending stiffness, and simulator's timescale~\cite{HoudiniVellum}, our method aims to predict vertex positions of a 3D garment mesh sequence $G^{\{1,...,t\}} \in \mathbb{R}^{t\times V \times 3}$. We assume that the garment mesh has a fixed topology. In this section, we first describe our use of virtual bones, how they drive the motion network and how we handle the variations in simulation parameters.




The virtual bones correspond to a set of bones $\{B\}$ that drive the garment deformations using rigid transformations, each of which controls a part of the garment during animation. Compared with 3D coordinates of vertices, the rigid transformations of virtual bones offer powerful and robust representations of complex deformations. In the animation, the deformations of the garment are driven by the virtual bones' motions using LBS: $G=LBS(M_v)$, where the virtual bones' motions $M_v$ consist of the rotations and translations of each of them $M_v^{\{1,...,t\}}=[R_{v,j}|T_{v,j}]^{\{1,...,t\}}$. The virtual bones are extracted from the smoothed simulation sequence $\hat{G}_{LF}^{\{1,...,t\}}$ using skin decomposition~\cite{BinhLe:TOG:2012, Liu:2020:HIS}, as is shown on the top row of Fig.~\ref{pipeline}. We highlight two examples of virtual bones and corresponding skin weights in Fig.~\ref{VirtualBoneWeights}.
The virtual bones are different from the skeletons of characters rigged by artists~\cite{LBS,kavan2009automatic}. First, artists tend to create hierarchical structures for bones to form skeletons when rigging characters, restricting the bones to rotating in relation to their parents. On the other hand, virtual bones are non-hierarchical and can be rotated and translated freely. Second, the artist-created skeletons have semantic meanings, while virtual bones are extracted from a simulation dataset and may not correspond to semantic representations. In its name, \textit{virtual} refers to lacking semantic meaning, and \textit{bone} means that they are rigid bones extracted by SSDR.


A motion network $D$ is a network driven by virtual bones whose motions $M_v$ are used to model the garment deformations caused by body motions $M_{b}$. Using the virtual bone as an intermediate representation, the motion network can effectively model the complex deformations on garments while retaining fine details. In our motion network, we use the low- and high-frequency modules $D_{LF},D_{HF}$ sequentially to estimate the corresponding low- and high-frequency garment deformations  $G_{LF},G_{HF}$, where the former refer to the overall garment shape and the latter represent the high-frequency details (fine wrinkles) on the garments. The low-frequency module $D_{LF}(M_b)$ transfers body motion sequence $M_{b}^{\{1,...,t\}}$ to virtual bones' motions $M_{v}^{\{1,...,t\}}$ using Gated Recurrent Units (GRU) \cite{GRU}. The low-frequency deformations can be generated by the skinning function of virtual bones: $G_{LF}=LBS(M_v)$. The high-frequency module $D_{HF}$ takes the virtual bones' motions estimated by the previous module and the local information on low-frequency mesh to predict high-frequency deformations $G_{HF}=D_{HF}(M_v, G_{LF})$. The former are processed using a GRU similar to the one in the low-frequency module, and the latter are extracted by a Graph Neural Network (GNN).

Our approach tries to estimate cloth deformations caused by the body motions and changes in the simulation parameters using separate learning methods. In particular, we model the deformations due to body motions using the motion networks; the variations in simulation parameters are modeled using an RBF kernel $\Psi$~\cite{RBF}. We use multiple motion networks $D_{1,...,K}$ as pivot motion networks, where each corresponds to a set of simulation parameters $\boldsymbol{\theta}_1,...,\boldsymbol{\theta}_K$. The garment mesh corresponding to the input simulation parameters $\boldsymbol{\theta}$ is a weighted summation of resultant meshes of pivot motion networks $G=\sum w_i G_i$, where the weighting coefficients depend on an RBF kernel that quantifies the difference between their simulation parameters and the input simulation parameters: $w_i=\Psi(\boldsymbol{\theta}_i,\boldsymbol{\theta})$. These are highlighted in Fig.~\ref{pipeline}.

\section{Learning Algorithm}


\subsection{Virtual Bone Generation}

\label{section:VirtualBone}

In this subsection we describe how to generate the virtual bones given the simulation sequence of garments $\hat{G}^{\{1,2,...,t\}}$. Similar to TailorNet~\cite{patel20tailornet}, our method decomposes the garment meshes to the low-frequency and the high-frequency deformations. Given a garment mesh $\hat{G}$, we perform Laplacian smoothing to obtain a smoothed mesh, which can be treated as the low-frequency deformation $\hat{G}_{LF}$, and take the residual as the high-frequency deformation $\hat{G}_{HF}=\hat{G}-\hat{G}_{LF}$.

We adopt the state-of-the-art skinning decomposition method Smooth Skinning Decomposition with Rigid Bones (SSDR)~\cite{BinhLe:TOG:2012} to transfer the low-frequency mesh sequence into an LBS model with a sequence of virtual bones' motions. Formally, given the mesh sequence $\hat{G}^{\{1,...,t\}}_{LF}$, the SSDR algorithm computes a rest pose mesh $P$, a skin weights matrix $W_{skin} \in \mathbb{R}^{V\times |B|} $, and a sequence of virtual bones' transformation matrices $\hat{M}_v=[\hat{R}_v|\hat{T}_v]=\{[\hat{R}_{v,j}^t|\hat{T}_{v,j}^t]\}$. The position of the mesh's $i$-th vertex at the $t$-th frame $v_i^t$ can be calculated as follows:
\begin{equation}
  v_i^t=LBS(\hat{M}_v^t;P,W_{skin})=\sum_{j=1}^{B}w_{skin,ij}(\hat{R}_{v,j}^tp_i+\hat{T}_{v,j}^t)+\epsilon,
\end{equation}
where $\epsilon$ is the residual of the SSDR algorithm, $p_i$ is the position of the $i$-th vertex of the rest pose mesh $P$, $w_{skin,ij}$ indicates the skin weight of $v_i$ regarding to the $j$-th virtual bone, and $\hat{R}_{v}, \hat{T}_{v}$ are the virtual bones' ground truth rotations and translations, respectively.


\subsection{Low-Frequency Module}

\label{section:LowFreq}

In the first stage of inference, the low-frequency module $D_{LF}$ transfers the body motion sequence to corresponding virtual bones' motions. In contrast to tight clothes, the history-dependent deformations play a central role in the loose-fitting garments' deformations. Therefore, we use recurrent neural networks~\cite{RNN} in the low-frequency module, which sequentially processes body poses by memorizing the current states and updating them according to the new poses. We specifically leverage GRU~\cite{GRU} as the building block, which has proven to be successful in modeling dynamic garments~\cite{santesteban19}. Since the translation of the body strongly affects the dynamics of loose-fitting garments, for each frame, we concatenate the body translation with the rotation of body joints to construct the input of the low-frequency module $M_b^{t} =(T_b^t||R^t_{b_{1}}||...||R^t_{b_{K}})$, where $||$ represents the concatenation operator and $K$ is the number of body joints. The outputs of the module are the rotations and translations of all virtual bones corresponding to this frame:
$M_v^t=(T^t_{v,{1}}||...||T^t_{v,{|B|}})||(R^t_{v,{1}}||...||R^t_{v,{|B|}})$. The network is trained via the loss containing two terms:
\begin{equation}
  \begin{split}
    \mathcal{L}_{LF}^t = & ||\hat{G}_{LF}^t-LBS({M}^t_v;P,W_{skin})||_2 + \\ & \lambda_{Lap} ||\Delta(\hat{G}_{LF}^t)-\Delta(LBS({M}^t_v;P,W_{skin}))||_2,
  \end{split}
\end{equation}
where the first term pushes the mesh generated by LBS to be close to the ground truth low-frequency mesh, and the second term encourages the Laplacians between the two meshes to be similar. $\lambda_{Lap}$ is a predefined weight factor. $P$ and $W_{skin}$ are precomputed by the skin decomposition algorithm and fixed during network training.


\subsection{High-Frequency Module}
\label{section:HighFreq}

In this subsection, we introduce the high-frequency module $D_{HF}$, which estimates the high-frequency deformation $G_{HF}$ of the garment. Our approach is based on the strategy used by GNNs~\cite{EdgeConv,HeterSkinNet}, which leverages global features and local features to enhance the size of the receptive field. This module treats the virtual bones' motions as global information, extracts the local information from the low-frequency mesh and leverages them to estimate the high-frequency deformations. We also note that we choose virtual bones' motions rather than body motions as the global information, as the former better represents the deformations of the garments.

To extract the local information of the vertices, we first generate the low-frequency mesh, which is the LBS-deformed mesh driven by virtual bones' motions $G_{LF}=LBS({M}^t_v;P,W_{skin})$, and use a stack of graph convolutional layers to extract the information from it. The input and output for every vertex are its position $p_{v_i}\in \mathbb{R}^3$ and its corresponding local feature $f^t_{local,v_i}$, respectively. Specifically, we adopt EdgeConv operator~\cite{EdgeConv}, which has been shown to be efficient for extracting the local features from point clouds~\cite{EdgeConv} and meshes~\cite{RigNet,HeterSkinNet}. In the global stream, the overall motion information is processed using a GRU structure similar to the low-frequency module, while changing the input to the motions of virtual bones $M_v$ and the output to the global feature of every vertex $f_{global,v_i^t}$. The high-frequency deformation of each vertex is obtained by processing the concatenation of local and global features via a shared weight MLP. The result of the motion network is a simple addition of low- and high-frequency deformations $G=G_{LF}+G_{HF}$.




For the loss function, besides position term, we employ a collision term to avoid body-garment collisions: $\mathcal{L}_{collision}^t =max(-n_{B,k}^t(v_i^t-v_{B,k}^t),0)$, where $v_{B,k}^t, n_{B,k}^t$ are position and normal of the $k$-th body vertex, which is closest to the $i$-th estimated garment vertex.

\subsection{RBF-Based Simulation Parameter Variation}
\label{section:RBF}

We describe our method combining different motion networks corresponding to different sets of simulation parameters. To ease the task of modeling garment deformations under different poses and simulation parameters, our method disentangles the deformations caused by the two sources, and models them using different networks.

Specifically, we use a convex combination of the pivot motion networks with different simulation parameters $D_{\boldsymbol{\theta}_{1,...,K}}(M_b)$ to generalize to garment deformations under unseen simulation parameters $\boldsymbol{\theta}$. To build correspondence between pivot networks and simulation parameters, we train each of them using animation sequences simulated with corresponding simulation parameters. The pivot motion networks are chosen from the training set. The process for choosing them is explained below. The calculation of mixture weights is based on Radial Basis Function (RBF)~\cite{RBF}, where motion networks with similar simulation parameters have higher weights and others lower. The overall computation corresponds to:
\begin{equation}
  G=\sum_{i=0}^{K}\overline{\Psi}(\boldsymbol{\theta}_i,\boldsymbol{\theta})D_{\boldsymbol{\theta}_i}(M_b),
\end{equation}
\begin{equation}
  \Psi(\boldsymbol{\theta}_i,\boldsymbol{\theta})=\exp(-\frac{||g(\boldsymbol{\theta}_i)-g(\boldsymbol{\theta})||^2_2}{2 \sigma^2}),
\end{equation}
where $g$ is an MLP that projects the simulation parameters to the latent space, $\Psi$ is an RBF kernel with bandwidth $\sigma$ evaluating mixture weights based on the distance between simulation parameters' latents, and $\overline{\Psi}$ is the normalized mixture weights.

To choose the $K$ pivot motion networks that best cover the space of different simulation parameters, we first select the networks having with least one highest or lowest simulation parameter among all networks as pivot motion networks to fit each of the other networks. Next, we iteratively fit non-pivot networks using the pivot motion networks and add the network with the highest RMSE to pivot motion networks until we get $K$ pivot motion networks.

\section{Results and Experiments}

We use a garment animation dataset consisting of three different types of garments driven by a digital avatar.  To animate the character, we collect the dance motions of MikuMikuDance (MMD) videos~\cite{wiki:MMD} from the internet, which consist of complex body motions. To prepare a dataset with different simulation parameters, we choose three simulation parameters that greatly affect the simulation results of garments: bending stiffness, mass density, and time scale. We sample 10 sets of simulation parameters and split 8/2 sets as training/testing sets. For each set, we generate 40000 frames of simulation data using the Houdini Vellum simulation system~\cite{HoudiniVellum}. Our networks are implemented using PyTorch~\cite{PyTorch} and PyTorch Geometric~\cite{PyTorchGeometric}. In the low-frequency module $D_{LF}$, the network is a single GRU layer followed by a linear layer. In the high-frequency module $D_{HF}$, we use a similar structure in the GRU and stack three EdgeConv layers for GNN. In our setting, we use 80 virtual bones and 8 pivot motion models. More details about our dataset, network architectures, hyper-parameters and other implementation details are presented in the supplementary material.


\subsection{Results and Evaluation}

\begin{figure}[htbp]
  \centering
  \includegraphics[width=\linewidth]{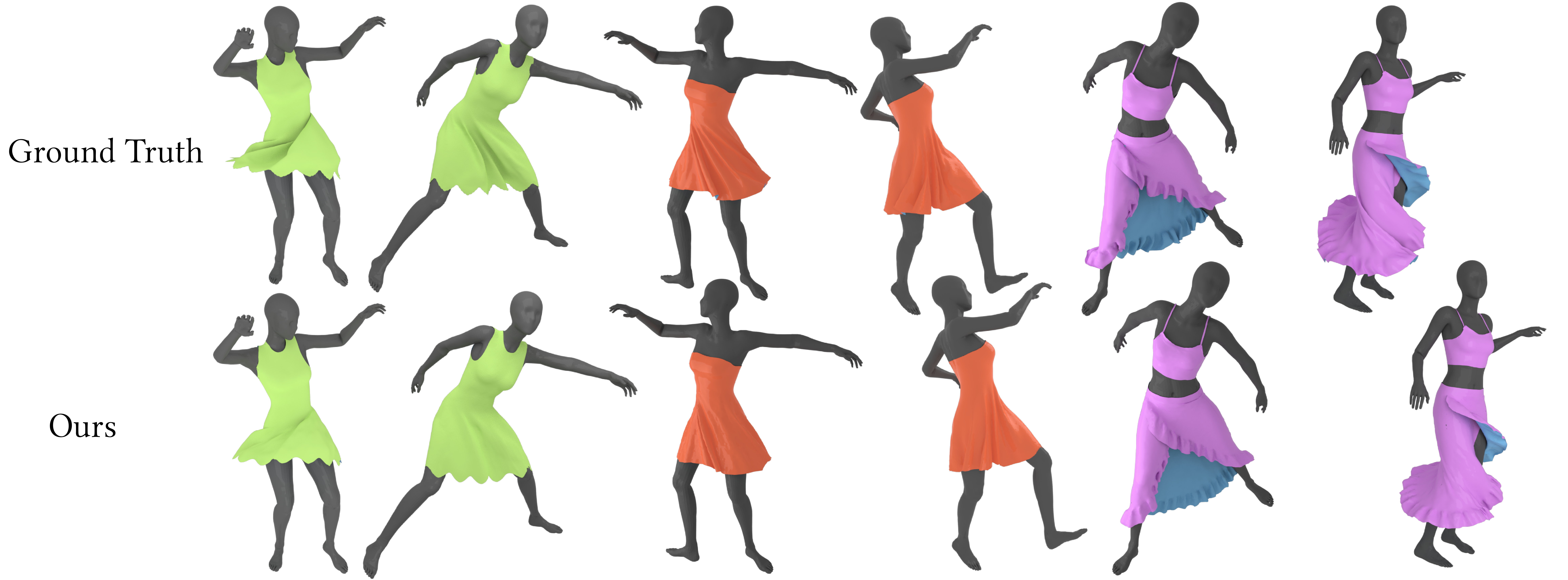}
  \caption{Given unseen body motions, our method can estimate garment meshes close to ground truth meshes for different types of garments.}
  \label{UnseenMotions}
\end{figure}

We present the result of our network for unseen motions in Fig.~\ref{UnseenMotions}. Our method supports complex deformations of garments under large quick motions, e.g., legs wide apart or swirling swiftly. The results are close to the ground truth in the overall shape but may lack some fine details or wrinkles.

\begin{figure}[htbp]
  \centering
  \includegraphics[width=\linewidth]{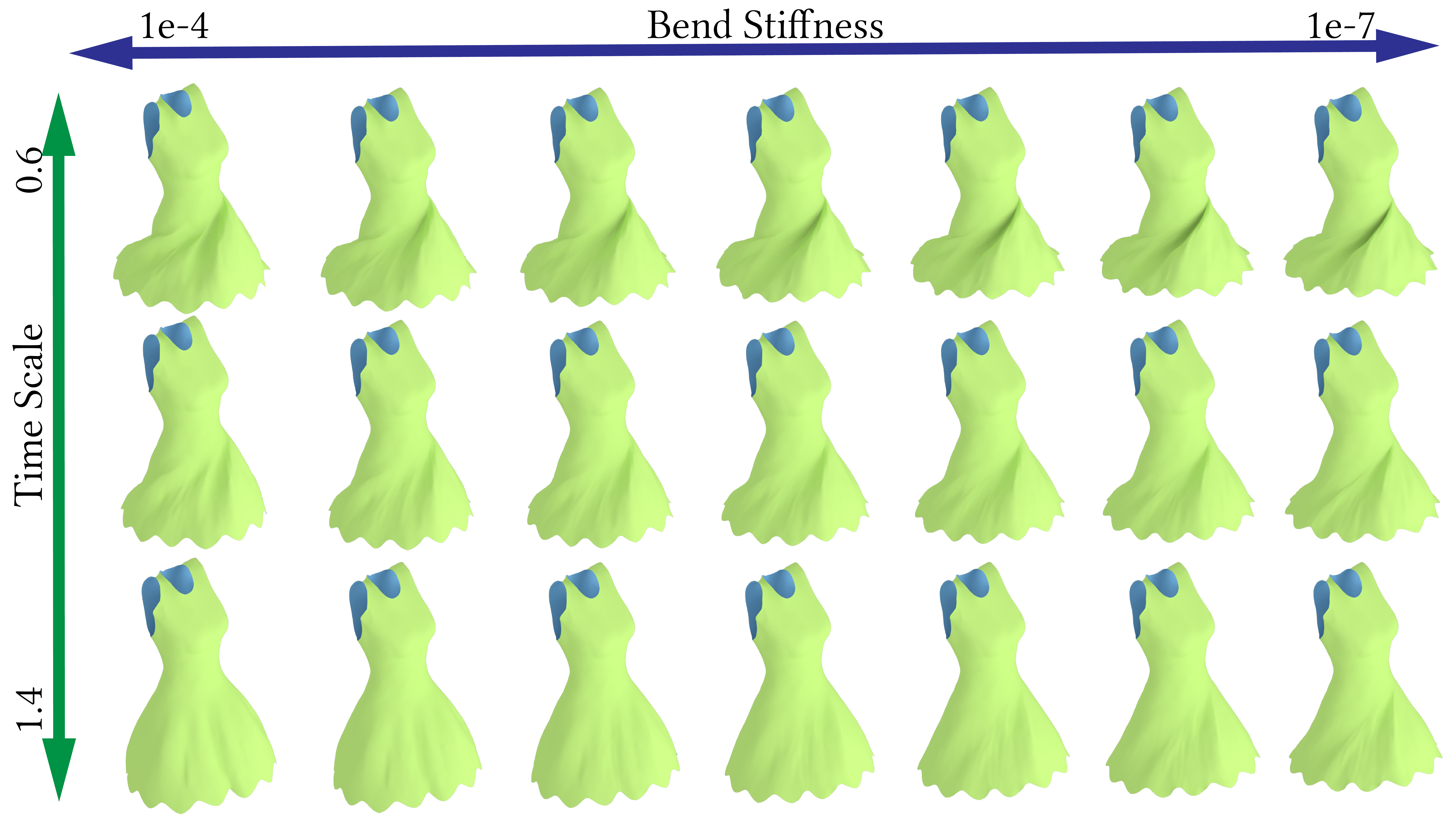}
  \caption{Estimation of the same frame under different simulation parameters (bending stiffness and timescale). Our method effectively predicts garment meshes corresponding to different simulation parameters in real time.}
  \label{DifferentSimParams}
\end{figure}

We present the performance of our network under different unseen simulation parameters in Fig.~\ref{DifferentSimParams}. Our network is capable of handling different simulation parameters: from left to right, the bending stiffness of the garment lessens, which leads to deeper creases; from top to bottom, the timescale of the simulator increases, causing the garments to be closer to their rest states. Tuning simulation parameters in physical simulations is tedious and consumes a lot of time when generating the results even with proper parameters (1.5 fps for garments used in our benchmarks). Other data-driven methods, which require regeneration of the dataset and training (about 12 hours), are more expensive. Our method supports interactively adjusting simulation parameters and generating new deformations.

\begin{figure}[htbp]
  \centering
  \includegraphics[width=\linewidth]{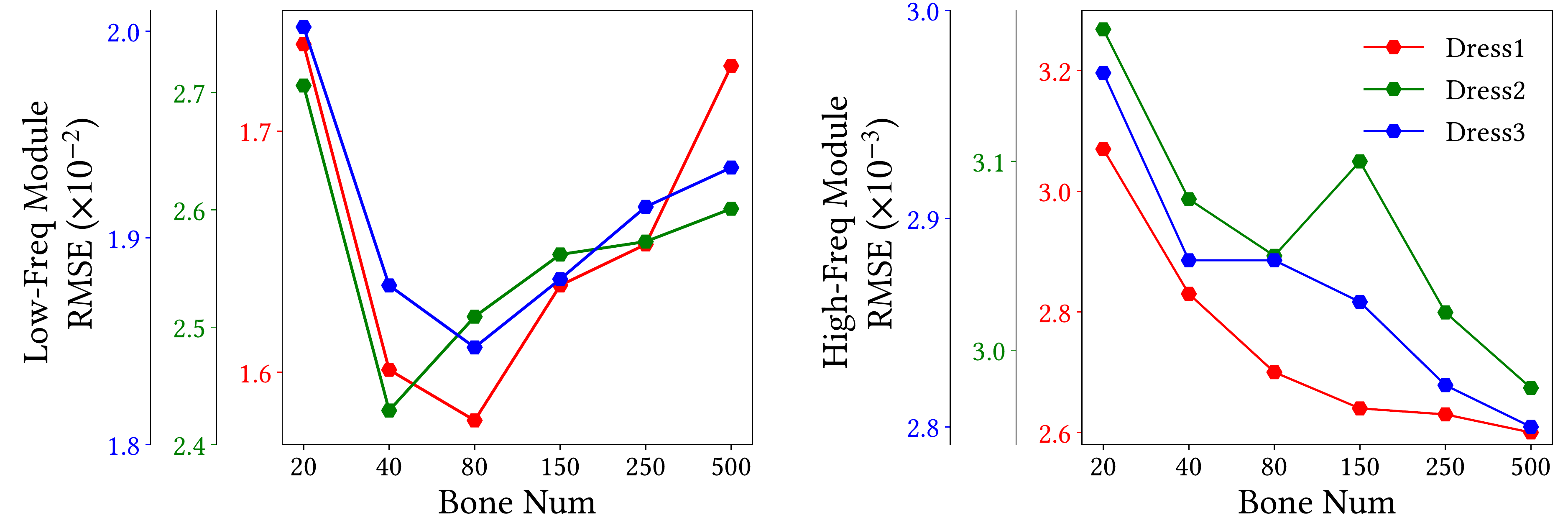}
  \caption{Comparison of low- and high-frequency modules' performances with different numbers of virtual bones. The low-frequency module estimates meshes with the lowest RMSE with 40/80 virtual bones, and the high-frequency module achieves the lowest RMSE with 500 virtual bones.}
  \label{HowManyBones}
\end{figure}

An important parameter in our pipeline is the number of garments' virtual bones. We evaluate the effects of this parameter by training low- and high-frequency modules with different numbers of virtual bones (20/40/80/150/250/500) for each garment. The quantitative results are shown in Fig.~\ref{HowManyBones}. In terms of the low-frequency module, our method achieves lower RMSE using 40/80 virtual bones for different garments. This may be attributed to the expressiveness of LBS and network capacity. When the number of bones is low, the LBS cannot fully capture the mesh, as the reconstruction error of SSDR for Dress1 is $1e-2$ with 20 virtual bones. When the number is high, the network capacity is limited. As for the high-frequency module, the network achieves lower errors with a higher number of virtual bones, because they carry more information about the mesh deformations. In the rest of our benchmarks, we use 80 virtual bones for each garment for lower overall RMSE.

\subsection{Comparisons with Prior Methods}

\begin{table}[htbp]
  \caption{ We compare the characteristics of our method with prior data-driven methods. Some of these methods are limited to tight-fitting garments or static poses and may not explicitly support changing the simulation parameters. Methods considering the dynamics of garments estimate the their deformations using the body motions of previous frames.}


  \setlength{\tabcolsep}{0.8mm}
  {
    \begin{tabular}{c|ccc}
      Method                                       & \tabincell{c}{Loose-                   \\ fitting}  & Dynamic & \tabincell{c}{Sim-Param \\ Variations}  \\
      \midrule[1.2pt]
      \tabincell{c}{TailorNet \cite{patel20tailornet}                                       \\ GarNet \cite{garnet} \\ \cite{jin2020pixel} \\\cite{deepsd} \\\cite{FullyConvolutionalGNN} \\ \cite{corona2021smplicit}\\ \cite{nannan}} & \xmark & \xmark & \xmark \\
      \hline
      \tabincell{c}{ Deepwrinkles \cite{deepwrinkles}                                       \\ \cite{santesteban19,santesteban21}  } & \xmark & \cmark & \xmark \\
      \hline
      \tabincell{c}{ Intrinsic \cite{Intrinsic}                                             \\ DNG \cite{zhang2021dynamic} \\ \cite{NearExhaustive}} & \cmark & \cmark & \xmark \\
      \hline
      \tabincell{c}{PBNS \cite{bertiche2020pbns} } & \xmark               & \xmark & \cmark \\
      \hline
      \tabincell{c}{N-Cloth \cite{Ncloth} }        & \cmark               & \xmark & \xmark \\
      \hline
      Ours                                         & \cmark               & \cmark & \cmark \\
    \end{tabular}
  }
  \label{characteristics}
\end{table}

In Table 1, we list the characteristics of different learning-based methods for modeling garments. Prior methods model tight garments~\cite{patel20tailornet,deepwrinkles,jin2020pixel,santesteban19,santesteban21,3dpeople,FullyConvolutionalGNN,cloth3d,garnet,bertiche2020pbns,deepsd} by rigging the garments using body motions, which causes ripping and non-smoothness on loose-fitting garments under large body motions. Moreover, the static methods~\cite{patel20tailornet,jin2020pixel,3dpeople,FullyConvolutionalGNN,cloth3d,garnet,bertiche2020pbns,deepsd} estimate the garment using only the body pose of the current frame and may not capture all deformations. Other methods control the simulation parameters~\cite{bertiche2020pbns} using the weight of loss terms in the network, which are hard to tune and may have no direct correspondence with the simulator. Our method explicitly models simulation parameters corresponding to the simulator using an RBF kernel.

We compare our method with others and perform ablation studies on three components: the low- and high-frequency modules and the overall performance of the motion network. Since prior methods do not focus on variations in simulation parameters, we perform comparisons using the dataset simulated with one set of simulation parameters (bending stiffness 1e-7, density 0.04, and simulator's timescale 1.0). All methods are trained with the same motion set and evaluated using the same unseen motions. We use three metrics to quantitatively evaluate the estimated meshes, namely RMSE (Root Mean Squared Error), Hausdorff distance and STED (Spatio-Temporal Edge Difference)~\cite{STED}. The first two calculate distance between estimated and ground truth meshes, while STED measures difference between their dynamics. We demonstrate improvements on each module in Table~\ref{motion_module}. We have also compared our approach with other methods targeting overall deformations.

We also tested our method on public datasets containing tight garments. We provide the quantitative results in Table \ref{public_dataset}. The quantitative results show that our method generates comparable results with TailorNet \cite{patel20tailornet} and \cite{santesteban19}.

We evaluate the choice of splitting the estimation of low- and high-frequency deformations by quantitatively comparing the performance of our network with other networks that have structure similar to our low-frequency module.
We present the quantitative results in the left column of Table~\ref{motion_module}., which shows that separating the prediction slightly enhances the performance. This is due to the fact that our high-frequency module leverages the prediction results of the low-frequency module, as shown in the ablation study of the high-frequency module.

\begin{figure}[htbp]
  \centering
  \includegraphics[width=\linewidth]{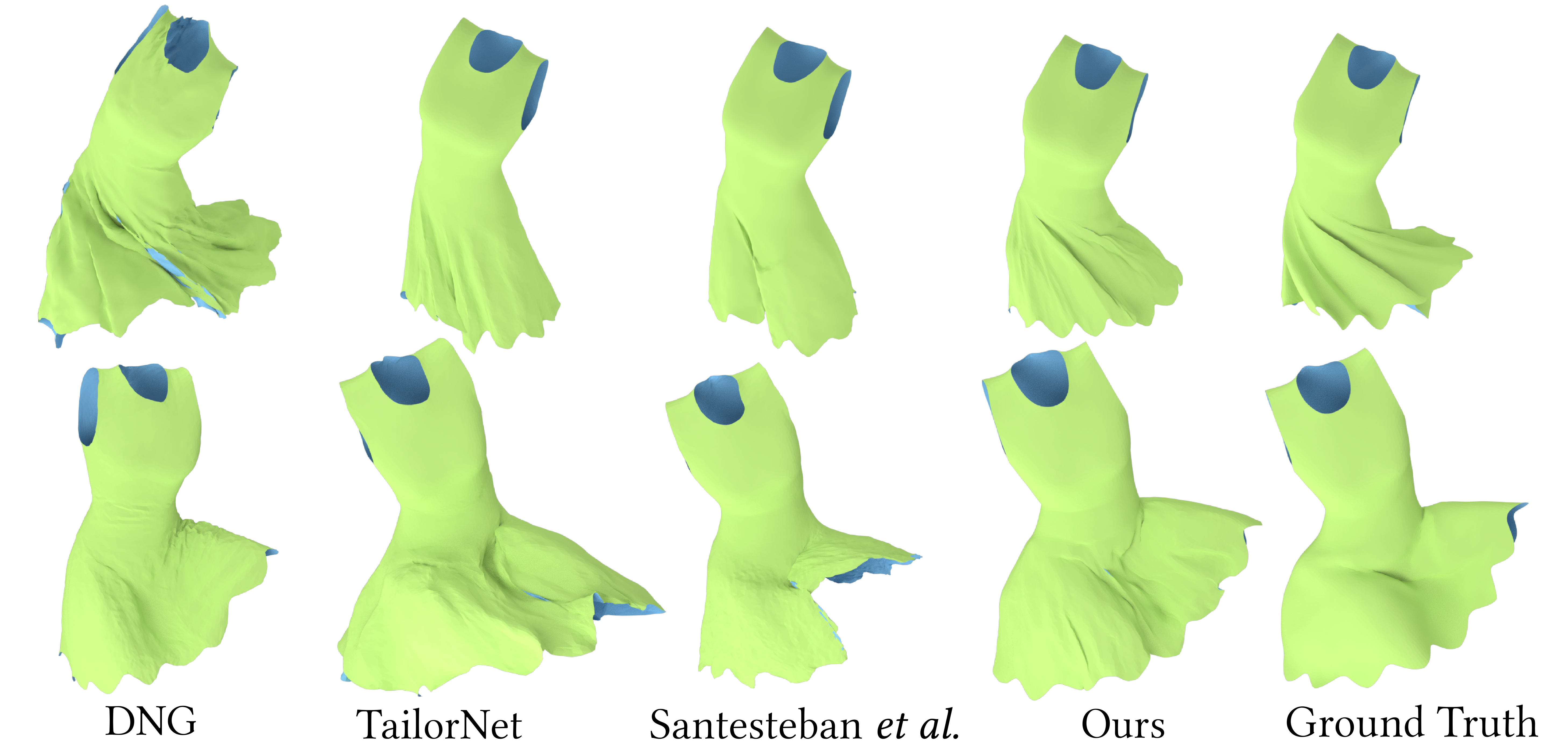}
  \caption{Qualitative comparisons of the low-frequency module with different methods. Top / bottom row: Dress 1 driven by avatars swiftly spinning and with legs wide apart. Our method tends to generate the 3D deforming mesh with the highest quality that is closest to the ground truth. }
  \label{Qualitative-low-freq}
\end{figure}

For the low-frequency module, we compare our method with TailorNet~\cite{patel20tailornet}, Dynamic Neural Garments (DNG)~\cite{zhang2021dynamic}, and \cite{santesteban19}. For DNG, we only use their method of generating coarse meshes. In the middle of Table~\ref{motion_module}, we provide a quantitative comparison. We observe that our method outperforms the baseline methods for all metrics, where RMSE is about $20\%$ lower and Hausdorff distance and STED are about $10\%$ lower than the best of them. We present two detailed comparisons in Fig.~\ref{Qualitative-low-freq}, where our method tends to generate a 3D mesh with the highest quality that is closest to the ground truth. Methods using static poses such as TailorNet~\cite{patel20tailornet} fail to predict the loose part of the dress, and our improvement comes from both virtual bones and modelling dynamics. \cite{santesteban19} generate meshes with high distortions on parts between thighs when the avatar's legs are wide open, as the garment is rigged to closely follow the body. Moreover, DNG ~\cite{zhang2021dynamic} may generate results with noticeable artifacts on loose parts.

\begin{figure}[htbp]
  \centering
  \includegraphics[width=\linewidth]{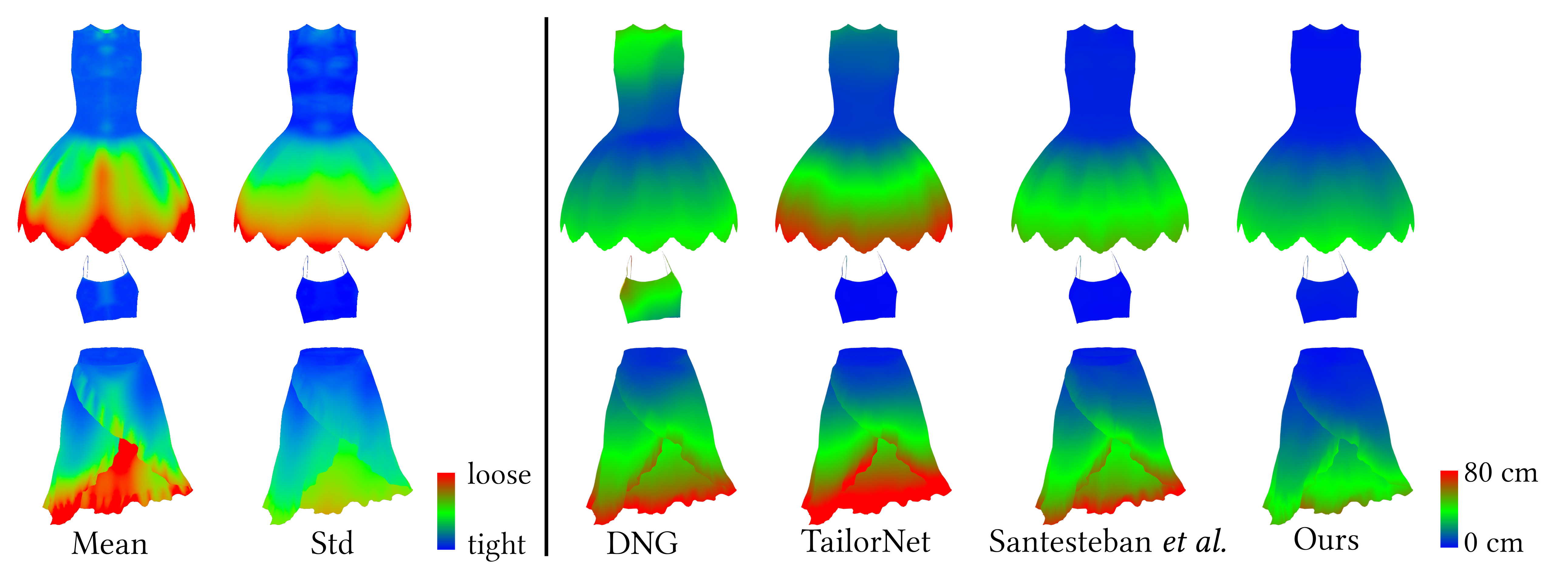}
  \caption{We highlight loose parts on the garment and per vertex mean errors. On the left, we render the vertices' mean and std distances from the nearest body vertices during animation, where the red color indicates loose parts of the garment. On the right, we compare the per vertex mean error of our method with others, where ours results in lower errors on loose parts.}
  \label{loose}
\end{figure}

\begin{table*}[htbp]
  \caption{Comparison with other methods and ablations for our motion model.}
  \setlength{\tabcolsep}{1.2mm}
  {

    \begin{tabular}{cc|cc|cccc|ccccc}
                              &                        & \multicolumn{2}{c|}{Overall} & \multicolumn{4}{c|}{Low-Frequency Module} & \multicolumn{4}{c}{High-Frequency Module}                                                                                                                         \\

                              &                        & \tabincell{c}{Ours                                                                                                                                                                                                                           \\ \scriptsize (w/o splitting \\ \scriptsize Low-High Freq)} & Ours & \tabincell{c}{TailorNet }   & \tabincell{c}{DNG}       & \tabincell{c}{\small [Santesteban \\ et al. 2019]}     & Ours   & \tabincell{c}{TailorNet }         & \tabincell{c}{\small [Chen et \\ al.  2021]}     & \tabincell{c}{Ours \\ \scriptsize (using body \\ \scriptsize motions)}   & \tabincell{c}{Ours \\ \scriptsize (w/o local \\ \scriptsize stream)}      & Ours  \\
      \midrule[1.2pt]
      \multirow{3}{*}{Dress1} & RMSE $\downarrow$      & 17.62                        & \textbf{17.38}                            & 32.94                                     & 22.78  & 20.58  & \textbf{15.76}  & 6.20   & 3.60   & \centering 3.07  & \centering 2.70           & \textbf{2.64}    \\
                              & Hausdorff $\downarrow$ & 70.94                        & \textbf{69.07}                            & 91.79                                     & 75.23  & 75.62  & \textbf{65.52}  & 16.28  & 16.02  & \centering 15.89 & \centering 13.31          & \textbf{13.21}   \\
                              & STED   $\downarrow$    & 0.0810                       & \textbf{0.0786}                           & 0.0977                                    & 0.0955 & 0.0871 & \textbf{0.0729} & 0.0346 & 0.0155 & 0.0151           & 0.0139                    & \textbf{0.0137}  \\
      \hline
      \multirow{3}{*}{Dress2} & RMSE  $\downarrow$     & 27.35                        & \textbf{27.10}                            & 40.48                                     & 37.17  & 30.04  & \textbf{24.29}  & 5.37   & 4.91   & \centering 3.46  & \centering 3.11           & \textbf{3.05}    \\
                              & Hausdorff $\downarrow$ & 100.65                       & \textbf{98.93}                            & 128.56                                    & 116.63 & 104.15 & \textbf{92.31}  & 20.01  & 22.30  & \centering 19.94 & \centering 18.21          & \textbf{18.01}   \\
                              & STED   $\downarrow$    & 0.0782                       & \textbf{0.0732}                           & 0.0905                                    & 0.1046 & 0.0759 & \textbf{0.0691} & 0.0300 & 0.0168 & 0.0151           & 0.0144                    & \textbf{0.0142}  \\
      \hline
      \multirow{3}{*}{Dress3} & RMSE $\downarrow$      & 22.01                        & \textbf{21.93}                            & 35.70                                     & 23.55  & 23.59  & \textbf{18.47}  & 5.16   & 3.97   & \centering 3.15  & \centering 2.90           & \textbf{2.86}    \\
                              & Hausdorff $\downarrow$ & 62.31                        & \textbf{62.30}                            & 89.88                                     & 71.78  & 75.40  & \textbf{58.76}  & 15.94  & 16.47  & \centering 15.49 & \centering \textbf{13.49} & 13.51            \\
                              & STED  $\downarrow$     & 0.0841                       & \textbf{0.0807}                           & 0.1045                                    & 0.0967 & 0.0850 & \textbf{0.0779} & 0.0383 & 0.0179 & 0.0167           & 0.01555                   & \textbf{0.01553} \\
    \end{tabular}
  }
  \label{motion_module}
\end{table*}

\begin{table}[htbp]
  \caption{Comparison on the overall performance of the motion network on the dataset of \cite{santesteban19} and TailorNet \cite{patel20tailornet}. T-shirt and skirt are from \cite{santesteban19} and pants are from TailorNet \cite{patel20tailornet}.}
  {
    \begin{tabular}{cc|cccc}
                               &                        & \tabincell{c}{TailorNet} & \tabincell{c}{[Santesteban                   \\ et al. 2019]}     & Ours     \\
      \midrule[1.2pt]

      \multirow{3}{*}{T-shirt} & RMSE $\downarrow$      & \textbf{9.90}            & 10.25                      & 10.52           \\
                               & Hausdorff $\downarrow$ & \textbf{27.02}           & 29.56                      & 31.51           \\
                               & STED   $\downarrow$    & \textbf{0.0418}          & 0.0449                     & 0.0452          \\
      \hline
      \multirow{3}{*}{Skirt}   & RMSE  $\downarrow$     & 22.95                    & 20.96                      & \textbf{19.91}  \\
                               & Hausdorff $\downarrow$ & \textbf{76.80}           & 87.01                      & 83.39           \\
                               & STED   $\downarrow$    & 0.0757                   & 0.0745                     & \textbf{0.0722} \\
      \hline
      \multirow{3}{*}{Pants}   & RMSE  $\downarrow$     & \textbf{4.84}            & 4.91                       & 5.08            \\
                               & Hausdorff $\downarrow$ & \textbf{14.46}           & 14.87                      & 18.75           \\
                               & STED   $\downarrow$    & \textbf{0.0127}          & 0.0129                     & 0.0166          \\
    \end{tabular}
  }
  \label{public_dataset}
\end{table}

\begin{figure}[htbp]
  \centering
  \includegraphics[width=\linewidth]{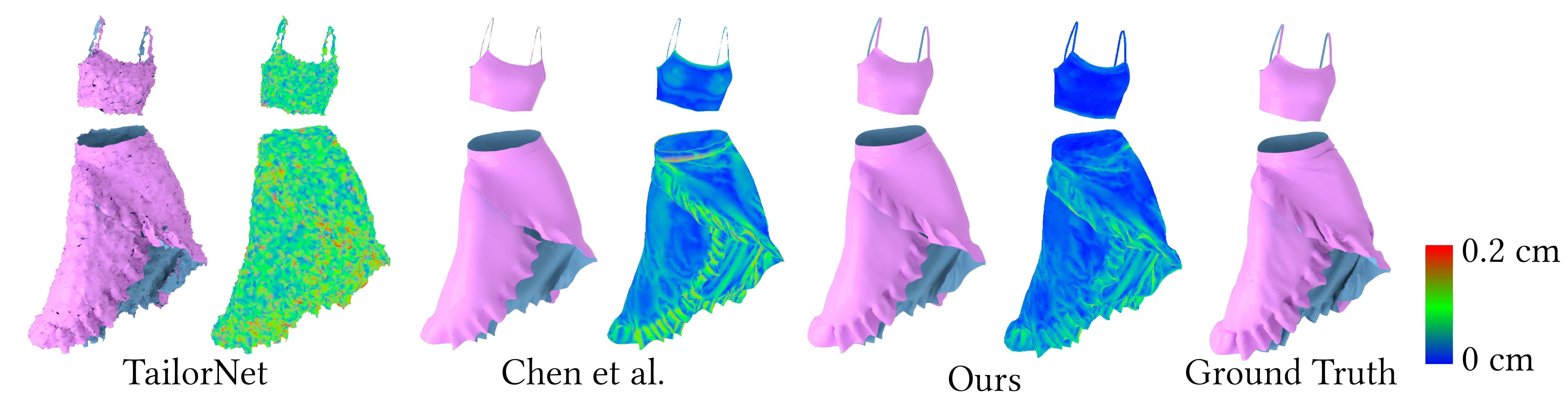}
  \caption{Qualitative comparison of the high-frequency module, the results and corresponding error maps for dress 2 driven by swift body motions. Our method estimates garment meshes with rich details and the lowest RMSE.}
  \label{Qualitative-high-freq}
\end{figure}

For the high-frequency module, we compare our method with \cite{patel20tailornet} and \cite{chen2021deep} and perform two ablation studies: the first uses the body motions as inputs, and the second does not use the local information extracted from the low-frequency mesh. We provide a quantitative comparison in the right column of Table~\ref{motion_module}, where our method outperforms the baseline methods. We present a detailed example in Fig.~\ref{Qualitative-high-freq}. Since TailorNet only takes static body motions as inputs, it generates non-smooth meshes when the body moves quickly. Our method generates smaller errors than \cite{chen2021deep}, since our high-frequency module combines the local information of the low-frequency mesh and virtual bones' motions as input. The two ablation studies validate the choices of using virtual bones' motions and combining local information.


For variations of simulation parameters, we compare our method with two other methods. The first uses the inverse simulation parameter distances as the summation weights instead of RBF, and the second combines the simulation parameters with motions in the two modules of the motion network. The quantitative results are shown in Table~\ref{sim_param_variations}, and they highlight the benefits of using an RBF kernel. We also note that the values in Tables~\ref{motion_module}. and~\ref{sim_param_variations}. are not comparable, as they use different test sets.


\begin{table}[]
  \caption{Ablation results on methods handling sim parameter variations.}
  \begin{tabular}{cc|ccp{20pt} p{20pt}c}
                            &                        & Linear & Large Net & Ours            \\
    \midrule[1.2pt]
    \multirow{3}{*}{Dress1} & RMSE $\downarrow$      & 24.82  & 20.74     & \textbf{16.69}  \\
                            & Hausdorff $\downarrow$ & 82.94  & 78.69     & \textbf{69.32}  \\
                            & STED $\downarrow$      & 0.1147 & 0.1087    & \textbf{0.0894} \\
    \hline
    \multirow{3}{*}{Dress2} & RMSE $\downarrow$      & 32.78  & 30.25     & \textbf{26.98}  \\
                            & Hausdorff $\downarrow$ & 113.92 & 108.37    & \textbf{99.76}  \\
                            & STED $\downarrow$      & 0.0910 & 0.0859    & \textbf{0.0801} \\
    \hline
    \multirow{3}{*}{Dress3} & RMSE $\downarrow$      & 22.45  & 21.40     & \textbf{19.82}  \\
                            & Hausdorff $\downarrow$ & 72.43  & 69.38     & \textbf{62.58}  \\
                            & STED $\downarrow$      & 0.0921 & 0.0923    & \textbf{0.0823} \\
  \end{tabular}
  \label{sim_param_variations}
\end{table}

\section{Conclusion, Limitations, and Future Work}

We have presented the first learning-based method to learn the complex deformations of loose-fitting garments via bone deformations. We have evaluated it on complex benchmarks and highlighted improved performance over prior learning-based methods. Compared with methods only using 3D coordinates, our virtual bone-driven approach can better predict complex deformations that are frequently observed in loose-fitting garments. Moreover, the virtual bones' motions used in our method can better guide the estimation of high-frequency details. Even for tight garments, our method generates results comparable with prior methods. Our method can also handle the variations of simulation parameters using an RBF-based model.

Our approach has some limitations. Our method uses LBS as skinning method for simplicity. However, its linear property may result in artifacts on areas influenced by multiple bones. Incorporating other skinning methods into our pipeline may be a good direction. The complex deformations can also result in self-collisions in the garment, which need to be prevented~\cite{tan2021lcollision}.






\bibliographystyle{ACM-Reference-Format}
\bibliography{ref}

\end{document}